 \documentclass[aps,twocolumn,pra,superscriptaddress,preprintnumbers,amsmath,showpacs,tightenlines]{revtex4}

\usepackage{epsfig,graphicx,times}

\def\<{\langle}
\def\>{\rangle}
\def\vac{|{\bf 0}\rangle}

\begin{document}

\title{Perfect function transfer and interference effects in interacting boson
lattices}

\author{Lian-Ao Wu} 
\affiliation{IKERBASQUE, Basque Foundation for Science, 48011,
Bilbao, Spain} \affiliation{Department of Theoretical Physics and
History of Science, The Basque Country University (EHU/UPV), 48080
Bilbao, Spain} \affiliation{Advanced Science Institute, The
Institute of Physical and Chemical Research (RIKEN), Wako-shi,
Saitama 351-0198, Japan}
\author{Adam Miranowicz} 
\affiliation{Advanced Science Institute, The Institute of Physical
and Chemical Research (RIKEN), Wako-shi, Saitama 351-0198, Japan}
\affiliation{CREST, Japan Science and Technology Agency (JST),
Kawaguchi, Saitama 332-0012, Japan} \affiliation{Faculty of
Physics, Adam Mickiewicz University, 61-614 Pozna\'n, Poland}
\author{XiangBin Wang}
\affiliation{Advanced Science Institute, The Institute of Physical and Chemical Research
(RIKEN), Wako-shi, Saitama 351-0198, Japan}
\affiliation{Department of Physics, Tsinghua University, Beijing 100084, China}
\author{Yu-xi Liu}
\affiliation{Advanced Science Institute, The Institute of Physical
and Chemical Research (RIKEN), Wako-shi, Saitama 351-0198, Japan}
\affiliation{CREST, Japan Science and Technology Agency (JST),
Kawaguchi, Saitama 332-0012, Japan} \affiliation{Institute of
Microelectronics and Tsinghua National Laboratory for Information
Science and Technology and Institute of Microelectronics, Tsinghua
University, Beijing 100084, China}
\author{Franco Nori}
\affiliation{Advanced Science Institute, The Institute of Physical
and Chemical Research (RIKEN), Wako-shi, Saitama 351-0198, Japan}
\affiliation{CREST, Japan Science and Technology Agency (JST),
Kawaguchi, Saitama 332-0012, Japan} \affiliation{Department of
Physics, Center for Theoretical Physics, Center for the Study of
Complex Systems, The University of Michigan, Ann Arbor, Michigan
48109-1120, USA}

\date{\today}

\begin{abstract}
We show how to perfectly transfer, without state initialization
and remote collaboration, arbitrary functions in interacting boson
lattices. We describe a possible implementation of state transfer
through bosonic atoms trapped in optical lattices or polaritons in
on-chip coupled cavities. Significantly, a family of Hamiltonians,
both linear and nonlinear, is found which are related to the
Bose-Hubbard model and that enable the perfect transfer of
arbitrary functions. It is shown that the state transfer between
two sites in two-dimensional lattices can result in quantum
interference due to the different numbers of intermediate sites in
different paths. The signature factor in nuclear physics can be
useful to characterize this quantum interference.
\end{abstract}


\maketitle \pagenumbering{arabic}

\section{Introduction}

Quantum information processing (QIP) often needs to transfer a
quantum state from one site to another \cite{Bennett}. For
example, in optical quantum communications, one may directly
transmit flying photons. However, in many other tasks, e.g.,
solid-state-based quantum computation, quantum state transmission
is not a trivial task. Completing such a task is often needed in
QIP, e.g., the quantum information exchange between two separate
processors. Therefore, it is very important to find physical
systems that provide robust quantum state transmission lines
linking different QIP processors.

In recent years, extensive investigations have been done on
quantum state transfer. In particular, many results have been
obtained on qubit-state transfer through spin chains with various
types of neighbor couplings. The original idea of quantum state
transfer through a system of interacting spins-1/2 was introduced
by Bose \cite{Bose}. Afterward, Chirstandl {\em et
al.}~\cite{Christandl} and independently Nikolopoulos {\em et
al.}~\cite{Nikolopoulos} found that perfect state transfer (PST)
is possible~\cite{Burgarth} in spin-1/2 networks without any
additional actions from senders and receivers, including not
requiring switching on and off qubit couplings. These results have
triggered much interest in demonstrating and optimizing reliable
state transfer in different models of interacting spin chains (see
e.g.,~\cite{Bose07,Stolze,Koji} and references therein). For
example, Di Franco {\em et al.}~\cite{DiFranco} clarified that PST
can be realized in spin-1/2 chains with neither state
initialization of the medium nor fine-tuned control pulses.
Recently, this scheme \cite{DiFranco} was further
simplified~\cite{Markiewicz} by not using remote collaboration of
senders and receivers. Thus, the PST of qubit states only requires
access to two spins at each end of the spin-1/2 chain.

\begin{figure}[ht]
\includegraphics[width=8.5 cm]{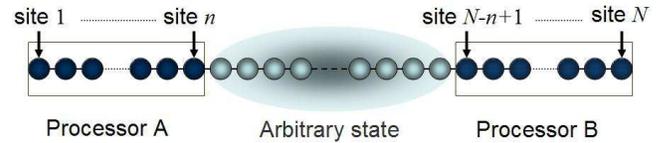}
\caption{(Color online) Our protocol enables a {\em perfect mirror
transfer} of any $n$-variable function $f\;$ from processor A to
processor B through a bosonic chain initially in an {\em
arbitrary} state.}
\end{figure}

Here, we show how to achieve PST of {\em any function} through
bosonic lattices. Earlier schemes often assume that all the spins
(besides the edge spins) are in the ground state. Here, we do not
assume this.

So far, only a few works~\cite{Plenio,Yung,Paz,Feder} investigated
the perfect transmission of an arbitrary {\em continuous}-variable
quantum state. As it is well known, many important quantum states
belong to this class, e.g., squeezed and coherent states. Also it
is possible to make QIP with such states for many important tasks.
Here, we propose a protocol to perfectly transfer an unknown
$n$-variable function from a processor at one end of a boson chain
to another processor at the other end (see Fig. 1). In the former
schemes, {\em exponential}-type states (e.g., coherent or squeezed
states) cannot be perfectly transferred through spin networks. It
is important to emphasize that our proposal {\em can} do this. By
using continuous-variable states, our protocol can be designed
directly, since there is a one-to-one correspondence between
continuous variables and bosonic operators.

The interaction required for perfect state transfer in our model
can be implemented using the Bose-Hubbard model---a paradigm for
studying correlated bosonic systems, such as optical lattices,
Josephson junction arrays, topological excitations, and so on
(see, e.g., \cite{Bruder} and references therein).

We stress that there is a family of Hamiltonians that can perform
the same tasks. We derive this family via the so-called {\em
dressing} transformations, which will be described below.
Moreover, we show that the state transfer between two sites in
two-dimensional lattices can produce quantum interference that can
be characterized by the signature factor used in nuclear theory
spectroscopy~\cite{Ring}.

\section{Bose-Hubbard model}

Let us now consider dynamics of bosons, in a system with $N$
sites, governed by the Bose-Hubbard
Hamiltonian~\cite{Bruder,Jaksch98,Hartmann08,Reslen09}:
\begin{equation}
H=-\sum_{k=1}^{N-1} J_{k}(b_{k}^{\dagger }b_{k+1}+b_{k+1}^{\dagger
}b_{k})+\sum_{k=1}^N \epsilon _{k}n_{k}+H_{U}, \label{BHH}
\end{equation}
where $n_{k}=b_{k}^{\dagger }b_{k}$ is the number operator for the
bosons located at the $k$th site, $b_{k}^{\dagger }$ ($b_{k}$) is
the bosonic creation (annihilation) operator. For simplicity, we
set $\hbar=1$ and, hereafter, we drop the lower and the upper
bounds of summation. Equation~(\ref{BHH}) describes hopping bosons
in the presence of the on-site repulsion, given by
\begin{equation}
 H_{U}=U\sum n_{k}(n_{k}-1).
\label{N1e}
\end{equation}
The hopping (tunneling) matrix element between nearest-neighbor
sites is given by
\begin{equation}
J_{k}=\int d^{3} \! \mathbf{r}\,w^{\ast
}(\mathbf{r-r}_{k})[T+V_{\text{lat}}
(\mathbf{r)]}w(\mathbf{r-r}_{k+1}), \label{BHH2}
\end{equation}
where $w(\mathbf{r-r}_{k})$ is a single-atom Wannier function at
lattice site $k$, $V_{\text{lat}}(\mathbf{r)}$ denotes the optical
lattice potential, and $T$ is the kinetic energy of a single atom.
The parameter
\begin{equation}
\epsilon _{k}=\int d^{3} \! \mathbf{r}\,V_{T} (\mathbf{r)}|
w(\mathbf{r-r}_{k}) | ^{2} \approx V_{T}(\mathbf{r}_{k}),
\label{N2e}
\end{equation}
where $V_{T}( \mathbf{r})$ characterizes an additional external
trapping potential. The parameters in an optical lattice are
controllable. For instance, the well depth, $V_{0}=\max |
V_{\text{lat}}(\mathbf{r)}|$, of the optical lattice can be tuned
in real time by changing the power of the lasers. The parameters
$V_{T}(\mathbf{r}_{k}\mathbf{)}$ and $V_{0}$ can be controlled to
obtain desired values of $\epsilon _{k}$ and $J_{k}$.

The model described by Eq.~(\ref{BHH}) can, e.g., be implemented
in a system of interacting polaritons in periodic arrays of
coupled optical resonators with strong atom-photon coupling
\cite{Hartmann06} (for a review see \cite{Hartmann08}).
Large-scale ($N>100$) arrays of ultrahigh-finesse ($Q \sim 1\times
10^6$) coupled photonic-crystal nanocavities have been realized
experimentally \cite{Notomi08} demonstrating a relatively long
photon lifetime ($\sim 1$ ns) with low propagation and coupling
losses. Also ultrahigh-finesse toroidal fiber-coupled
microresonators have been proposed \cite{Hartmann06}, for which
the observed coupling between fundamental modes of a fiber and a
microresonator much exceeds losses into other modes: specifically,
the experimental ideality factor was reported to be 99.97\%
\cite{Spillane}. Alternative implementations of Eq.~(\ref{BHH})
include: arrays of Josephson junctions (e.g.,
\cite{Geerligs,Zant}) and ultracold atoms trapped in optical
lattices (as suggested in \cite{Jaksch98} and first experimentally
realized in \cite{Greiner02}). For a review on the Bose-Hubbard
model and its physical implementations, see \cite{Bruder}. We note
that, in the latter systems, it is more difficult, in comparison
to coupled arrays of cavities, to access and to control individual
lattice sites due to the small distances between the sites
\cite{Hartmann08}.

We note that when the nonlinear term $H_{U}$ in Eq.~(\ref{BHH}) is
negligibly small, Eq.~(\ref{BHH}) becomes a linearly coupled
bosonic Hamiltonian, which is equivalent to that of the on-chip
coupled cavities (e.g., as theoretically described in
Refs.~\cite{Hartmann06},~\cite{Bliokh}, and~\cite{Zhou08}). The
controllable parameter $\epsilon _{k}$ of these cavities can be
realized by, e.g., tunable transmission line resonators~(e.g.,
Refs.~\cite{Sandberg} and~\cite{Johansson}), while the tunable
coupling $J_{k}$ between each pair of resonators can be realized
by superconducting quantum interference device couplers. Recent
progress in manufacturing ultrahigh-finesse microcavities on a
chip \cite{Armani,Kippenberg,Xia,Colombe} indicates the
experimental feasibility of state transfers in such systems with
current or soon-to-be-available technology.

\section{Angular momentum and engineered Bose-Hubbard model}

The angular momentum vector $\mathbf{L}$ is a single-particle
operator that can have either a fermionic or a bosonic
representation. The bosonic representation of $\mathbf{L}$ can be
written~\cite{Wybourne} as
\begin{equation}
\mathbf{L=}\sum_{m,m'} \< lm| \mathbf{L}| lm^{\prime } \>
A_{lm}^{\dagger }A_{lm^{\prime }},
\label{N3e}
\end{equation}
where $| lm\> $ are eigenstates of the total angular momentum
$\mathbf{L}^{2}$ and $L_{z}$. Here, we define bosonic creation
operators $A_{lm}^{\dagger }$ such that, under rotation, they map
or transfer among themselves as a rank-$l$ irreducible spherical
tensor operator.

Reference~\cite{Christandl} maps the indices of the site numbers
of a one-dimensional chain into the magnetic quantum number $m$ of
the total angular momentum $l$, such that $l=\frac{N-1}{2}$ and
$m=-\frac{N-1}{2}+k-1.$ For instance, the magnetic number for the
first site is $m=-l=-\frac{N-1}{2}.$ With this mapping, the rank
$l$ irreducible spherical tensor bosonic operator $A_{lm}^{\dagger
}$ corresponds to the bosonic operator $b_{k}^{\dagger }$ at site
$k=m+\frac{N+1}{2},$ which obeys the tensor transformation
\cite{Wybourne},
\begin{equation}
R(\Omega )A_{lm}^{\dagger }R^{-1}(\Omega )=\sum_{m^{\prime
}}D_{m^{\prime }m}^{l}(\Omega )A_{lm^{\prime }}^{\dagger },
\label{N4e}
\end{equation}
where $R(\Omega )$ is a rotation operator with Euler angles
$\Omega =(\alpha, \beta, \gamma )$, and
\begin{equation}
D_{m^{\prime }m}^{l}(\Omega )=\exp(-im^{\prime
}\alpha)d_{m^{\prime }m}^{l}(\beta )\exp(-im\gamma)
\label{N5e}
\end{equation}
given in terms of Wigner's $d$-matrix~\cite{Biedenharn},
$d_{m^{\prime }m}^{l}(\beta )$. The three components of the
angular momentum vector may be expressed by atomic creation and
annihilation operators as
\begin{eqnarray}
L_{x} &=&\sum C_{k}(b_{k}^{\dagger
}b_{k+1}+b_{k+1}^{\dagger }b_{k}), \notag \\
L_{y} &=&i\sum C_{k}(b_{k}^{\dagger
}b_{k+1}-b_{k+1}^{\dagger }b_{k}), \notag \\
L_{z} &=&\sum n_{k}\left[k- \textstyle{\frac{1}{2}}(N+1)\right],
\label{N11}
\end{eqnarray}
where $C_{k}=\frac{1}{2}\sqrt{k(N-k)}$. The term $\sum n_{k} $ in
$L_{z}$ is the total boson number and commutes with other
operators. If we engineer $J_{k}$ and $\epsilon _{k}$ in
Eq.~(\ref{BHH}) such that~\cite{Christandl} $J_{k}=J C_{k}$ and
$\epsilon _{k}=\epsilon \left(\frac{N+1}{2}-k\right),$ then its
time evolution becomes
\begin{equation}
U(t)=\exp[i(JL_{x}+\epsilon L_{z}-H_{U})t]. \label{U(t)}
\end{equation}
The on-site repulsion $H_{U}$ can be adjusted by varying the
strength $U$.

\section{Linear case}

Let us first consider the case when $J$ is so large that we can
ignore $H_{U}$, so the Bose-Hubbard model becomes linear and the
evolution operator $U(t)$ corresponds to a rotation operator
$R(\Omega)$. Additionally, by assuming that $\epsilon =0$, the
evolution is described by
\begin{eqnarray}
U^{\dagger }(t)A_{lm}^{\dagger }U(t) = \sum e^{i(\pi/2)(m^{\prime
}-m)}d_{m^{\prime }m}^{l}(Jt)A_{lm^{\prime }}^{\dagger }\,.
\end{eqnarray}
When $t_{0}=\pi /J,$ this expression reduces to a simple form:
\begin{equation}
U^{\dagger }(t_{0}) \; b_{i}^{\dagger } \;U(t_{0})=r\; b_{N-i+1}^{\dagger
}\, , \label{transfer}
\end{equation}
where the factor
\begin{eqnarray}
 r=\exp\left(-i\pi \frac{N-1}{2}\right)
\label{signature}
\end{eqnarray}
is analogous to the so-called {\em signature} in nuclear structure
theory~\cite{Ring}. Below, we will discuss the interference
induced by this signature factor. There exist observable effects
for different signatures in nuclear spectroscopy. For instance,
the lowest (so-called bandhead~\cite{Ring}) energy differs for the
two cases $r=-i$ and $r=i$, for the same spin $K=1/2$. Here, $r=1$
when the total number of sites $N=5,9,13,$ \ldots \;. In this case
\begin{equation}
U^{\dagger }(t_{0}) \; b_{i}^{\dagger } \; U(t_{0})=b_{N-i+1}^{\dagger }.
\label{transfer1}
\end{equation}
Moreover, these results are valid for a general linear Hamiltonian
$H_{l}= \mathbf{v}\cdot \mathbf{L}$, with constant vector
$\mathbf{v}=(-J,0,-\epsilon ).$

\section{Perfect state transfer for the linear case}

The following results can be applied to various linear optical or
atomic systems. Now, let us assume that a multi-variable function
$f$ is encoded in an $n$-site processor A, such that
$f(x_{1},...,x_{n})$ is mapped into the state $ f(b_{1}^{\dagger
},...,b_{n}^{\dagger })\vac $ as follows:
\begin{eqnarray}
f(x_{1},...,x_{n}) \longrightarrow f(b_{1}^{\dagger
},...,b_{n}^{\dagger })\vac, \label{map}
\end{eqnarray}
where $\vac=|0\>^{\otimes N}$. For instance, a function $\alpha
x_{1}^{2}+\beta x_{n}^{2}$, with unknown coefficients $\alpha $
and $\beta $, is mapped into the state $(\alpha b_{1}^{\dagger
}b_{1}^{\dagger }+\beta b_{n}^{\dagger }b_{n}^{\dagger })\vac.$
Thus, any general function $f$ can be perfectly transferred as
\begin{eqnarray}
&&\hspace{-1cm} U(t_{0}) \; f(b_{1}^{\dagger },...,b_{n}^{\dagger
}) \; \vac
\nonumber \\
&=&f(U(t_{0})b_{1}^{\dagger }U^{\dagger
}(t_{0}),...,U(t_{0})b_{n}^{\dagger }U^{\dagger }(t_{0}))\; \vac
\nonumber \\
&=&f(b_{N}^{\dagger },...,b_{N-n+1}^{\dagger }) \; \vac. \label{N5}
\end{eqnarray}
The function $f$ operates sequentially, from $1$ to $n$ in
processor A, and reads in the opposite order in processor B. We
emphasize that the central subset part of the chain (Fig. 1)
should not be necessarily in the ground state and can be in an
arbitrary state. However, for simplicity and without loss of
generality, here, we work with the state $\vac$.

\section{Perfect state transfer with on-site repulsion}

When the strength of $H_{U}$ is relatively strong and the total
number $\sum n_{k}$ is much smaller than the number, $N$, of
sites, the system tends to have at most one atom at each site
because of the gap caused by $H_{U}.$ Moreover, in the limit when
there is only one boson, we can still transfer a function
perfectly. In this case, an arbitrary state $| \phi \> $ of the
whole system can be annihilated by the on-site repulsion
Hamiltonian $H_{U}| \phi \> =0.$ Therefore, this case is the same
as the one in Eq.~(\ref{N5}). The transfer would be perfect if we
are able to prepare an initial state
\begin{eqnarray}
 |\phi \> _{1}=\alpha \vac +\beta b_{1}^{\dagger }\vac,
\label{initial}
\end{eqnarray}
where only the first site is occupied. Thus, the state can be
perfectly transferred to $ | \phi \> _{N}$ in the same way as in
the linear case. Generally, an arbitrary state,
\begin{equation}
|\phi\rangle=\alpha \vac +\sum_{k=1}^{n}\beta _{k}b_{k}^{\dagger
}\vac, \label{N6e}
\end{equation}
with one atom at an $n$-site processor A can also be transferred
perfectly.

\section{Generalization}

A time-independent arbitrary unitary transformation $W$ does not
change the commutation relations among angular momentum
components, meaning that $H_{l}=\mathbf{v}\cdot \mathbf{L}$ is
mapped into $H_{l}^{\prime }=\mathbf{v}\cdot \mathbf{L}^{\prime
}$, where the components of $\mathbf{L}^{\prime }$ satisfy the
same commutation relations. However, this map may introduce new
effects. Indeed, the whole family of Hamiltonians generated by an
arbitrary $W$ can transfer functions perfectly as in
Eq.~(\ref{transfer1}). For spin systems, this transformation $W$
corresponds to the so-called {\em dressed qubit}~\cite{Wu03}. The
bosonic Hilbert space is infinite-dimensional and this allows more
flexibility to use additional transformations, including
continuous-variable transformations (this is not possible in spin
chains). As an example, here, we now consider transformations $W$
such that they act on each site individually, namely, $W=
\prod_{k=1}^{N}W_{k}$. For instance, the Hamiltonian
$H_{l}=JL_{x}$ becomes
\begin{eqnarray} H_{l}^{\prime } =H_{l}+ \sum C_{k} \{
2|\beta | ^{2} -[\beta ^{\ast }(b_{k}+b_{k+1})+{\rm h.c.}]\}
\label{N6}
\end{eqnarray}
under the displacement operator $W_{k}=\exp (\beta b_{k}^{\dagger
} -\beta ^{\ast }b_{k})$. Here, we set $\beta _{k}^{\prime }$'s
equal to $\beta$ for all sites so that the {\em dressed coherent
state (dressed displaced vacuum)}
\begin{equation}
|\boldsymbol{\beta}\>\equiv |\beta\>^{\otimes N} 
=\prod \exp (\beta b_{k}^{\dagger } -\beta ^{\ast }b_{k})\vac
\label{N7e}
\end{equation}
is invariant under the transformation $U(t_{0})$. Any function $f$
is mirror transferred via the coherent state
$|\boldsymbol{\beta}\>$:
\begin{eqnarray} &&\hspace{-1.5cm}
f(b_{1}^{\dagger }-\beta ^{\ast },...,b_{n}^{\dagger }-\beta
^{\ast
}) \; |\boldsymbol{\beta}\> \notag \\
&\longrightarrow & f (b_{N}^{\dagger }-\beta ^{\ast
},...,b_{N-n+1}^{\dagger}-\beta ^{\ast }) \;
|\boldsymbol{\beta}\>. \label{coherent}
\end{eqnarray}
Another example of dressing transformation occurs via the one-mode
squeezing operator $W_{k}=\exp [(\xi /2)(b_{k}^{2}-b_{k}^{\dagger
2})],$ where the transferred Hamiltonian reads
\begin{eqnarray}
 H_{l}^{\prime }&=&H_{l} \cosh \xi + H_{s} \sinh \xi,
\label{N8e}
\end{eqnarray}
where
\begin{eqnarray}
 H_{s}&=&\sum C_{k}(b_{k}^{\dagger }b_{k+1}^{\dagger
 }+b_{k+1}b_{k}).
\label{N8ee}
\end{eqnarray}
This induces squeezed states. Any function $f$ can be perfectly
mirror transferred via the {\em dressed squeezed vacuum}
\begin{equation}
|\boldsymbol{\xi}\>\equiv |\xi\>^{\otimes N} =\prod \exp [(\xi
/2)(b_{k}^{2}-b_{k}^{\dagger 2})]\vac \label{N9y}
\end{equation}
in the same way as in Eq.~(\ref{coherent}).

We can also introduce another useful nonlinear term by setting
$\xi $ as an operator. If we bosonize $\xi =\xi _{0}(c+c^{\dagger
})$ and assume a small $\xi _{0},$ the Hamiltonian~(\ref{N8e})
becomes
\begin{equation}
H_{l}^{\prime }=H_{l}+\xi _{0}(c+c^{\dagger })H_{s}, \label{N10e}
\end{equation}
which describes a nonlinear down-conversion effect. If the boson
$\xi$ is coupled to a collective rotational system, $\xi =\xi
_{0}J_{z},$ the additional term in the Hamiltonian becomes $\sum
C_{k} J_{z}(b_{k}^{\dagger }b_{k+1}^{\dagger
}+b_{k+1}b_{k})=H_sJ_{z}.$

For generalizations, Refs.~\cite{Christandl} and~\cite{Yung} also
discussed the possibility of other Hamiltonians that might not be
directly related to $H_{l}$, but play the same role as $H_{l}.$
Reference \cite{Yung} shows that perfect state transfers are
associated with the spectra of these Hamiltonians. Here, it is
useful to point out that these Hamiltonians must not have the same
eigenspectra as $H_{l}$. These Hamiltonians can generate, via
arbitrary dressing transformations $W$, other families which are
not equivalent to the family of $H_{l}$.

\section{Interference effects due to different numbers of sites in each
path}

The phase or signature, given by Eq.~(\ref{signature}), with
values of $\pm 1$ and $\pm i$ forming a $Z_{4}$ cyclic group, is
analogous to the phase gate for qubits. For instance, an unknown
state $ \alpha |\mathbf{0}\rangle +\beta \, b_{1}^{\dagger } \,
|\mathbf{0}\rangle $ can be transferred into $\alpha
|\mathbf{0}\rangle +r \, \beta \; b_{N}^{\dagger } \, |\mathbf{
0}\rangle .$ When the total number of sites, $N$, is even, the
signature $r=\pm i.$ When $N$ is odd, $r=\pm 1$. For a given
signature $r$, this ``phase gate'' is applied {\em during} the
perfect function transfer processes.

\begin{figure}[ht]
\includegraphics[width=8 cm]{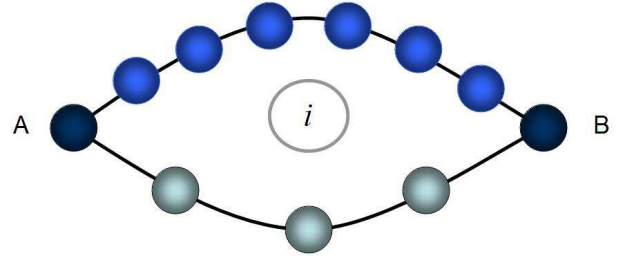}
\caption{(Color online) An example of site-number interference
effect for two paths containing $N=5$ sites (with signature factor
$r_1=1$) and $N=8$ sites (upper path with $r_2=i$). The
generalized ``optical paths'' in this case are the same (because
both paths are of equal length), but the number of sites, $N$, for
each path, is not. The overall phase factor $r$ for these two
paths is the product of the signature factors for each path. Thus
$r=r_1r_2=i$. Here, $r=i$ plays a role analogous to the
Aharonov-Bohm flux $\Phi$.}
\end{figure}

This phase gate induces an important effect. Consider now site A
as a sender and site B as a receiver. On a two-dimensional
lattice, we can consider different paths from site A to site B.
These paths can be designed such that the values of $N$ are {\em
different} for each path, but their ``generalized optical paths''
are the same. Different values of $N$ could result in up to four
different values of the signature factor $r$. A perfect function
transfer can simultaneously occur in, for instance, two paths.
This resembles the inference of two waves. States sent from site A
via different paths will interfere at site B (Fig. 2). Quantum
interference effects in directed paths can produce measurable
effects (see, e.g.,~\cite{Lin}).

Following Ref.~\cite{Jaksch98}, we can characterize the bosonic
field operators
\begin{equation}
\psi (\mathbf{r})= \sum_{k,p}b_{k}^{(p)} w(\mathbf{r-r}_{k}^{(p)})
\label{N12e}
\end{equation}
at time $t=0$, where $ p=1,2$ denotes the two paths with the
numbers of sites: $N_{1}$ and $N_{2}$. The field operator $\psi$
evolves as:
\begin{equation}
\psi (\mathbf{r,}t)=e^{-iJ(L_{x}^{(1)}+L_{x}^{(2)})t} \; \psi
(\mathbf{r}) \; e^{iJ(L_{x}^{(1)}+L_{x}^{(2)})t}, \label{N17a}
\end{equation}
where $L_{x}^{(1)}$ ($L_{x}^{(2)}$ ) is the angular momentum
operator of path 1 ($2$). At time $ t=t_{0}$,
\begin{eqnarray}
 \psi (\mathbf{r,}t_{0})&=&\sum_{k}[ b^{(1)}_{k}w(\mathbf{r-r}
^{(1)}_{N_{1}-k+1})
\nonumber \\
&&\quad + \; r \; b^{(2)}_{k}w(\mathbf{r-r}^{(2)}_{N_{2}-k+1})],
\label{N17b}
\end{eqnarray}
where we set $r=1$ for path 1. The average field intensity at
$\mathbf{r}$ is
\begin{eqnarray}
I(\mathbf{r},t\mathbf{)=}\left\langle \psi ^{\dagger }(
\mathbf{r},t) \; \psi (\mathbf{r},t)\right\rangle, \label{N15}
\end{eqnarray}
where $\left\langle ...\right\rangle $ denotes the expectation
value for the initial state. For simplicity, let us consider the
processor A to be localized at the origin $\mathbf{r}=0$. This
implies that $\left\langle (b^{(p')}_{k'})^{\dagger
}b^{(p)}_{k}\right\rangle =0,$ except for $k=k'=1.$ The initial
intensity is
\begin{equation}
I(\mathbf{r} ,0\mathbf{)}=4\left\langle (b^{(1)}_{1})^{\dagger }
b^{(1)}_{1}\right\rangle \vert w( \mathbf{r-r}_{1})\vert ^{2},
\label{N13e}
\end{equation}
where $\mathbf{r}_{1}\equiv \mathbf{r}_{1}^{(1)}=
\mathbf{r}_{1}^{(2)}$. Since we start from the first site
$\mathbf{r}=0$ at time $t=0,$ we approximately set
$b^{(1)}_{1}=b^{(2)}_{1}$ being the same boson. The final
intensity then becomes
\begin{eqnarray}
I(\mathbf{r},t_{0}\mathbf{)=}\left\langle (b^{(1)}_{1})^{\dagger }
b^{(1)}_{1}\right\rangle \left\vert w(\mathbf{r-r}_{N})\right\vert
^{2}(2+r+r^{\ast }) \label{N16}
\end{eqnarray}
because we set the two paths ending up at the same site
$\mathbf{r}_{N}\equiv\mathbf{r}^{(1)}
_{N_{1}}=\mathbf{r}^{(2)}_{N_{2}}.$ The intensity $I$ varies with
different values of the signature $r$. For example, the intensity
$I$ in Fig. 2 will be half of $I(\mathbf{r},0\mathbf{).}$ When the
signature $r=+1,-1,\pm i$, the factor $(2+r+r^{\ast })$ in
Eq.~(\ref{N16}) is equal to 4, 0 and 2, respectively. This
corresponds to constructive, destructive, and in-between
interferences. This site-number-dependent interference can be
generalized to include contributions from many directed
paths~\cite{Lin}. Note that in quantum interference path-integral
calculations~\cite{Lin}, summing over directed paths can require
summing over an enormously large number of different terms,
instead of just four different types of terms here. In other
words, the $Z_4$ group cyclic nature of the signature factor
reduces the huge number of different terms used for standard
interference calculations~\cite{Lin} to just summing over four
possible values of $r$, with $r=\pm1, \pm i$.

\section{Conclusions}

We have shown, by generalizing the recent results
of~\cite{DiFranco} and~\cite{Markiewicz}, that arbitrary functions
can be sent perfectly (without state initialization and remote
collaboration) through engineered interacting bosonic and qubit
chains. As an example, we have analyzed perfect state transfers
(using, e.g., ultracold bosonic atoms in optical lattices or
polaritons in coupled cavities) described by the Bose-Hubbard
model with properly designed site-dependent tunneling amplitudes
(the so-called Krawtchouk lattices).

In a more general case, we have studied a family of linear and
nonlinear Hamiltonians that enable perfect state transfers
according to dressing transformations leading to, e.g., dressed
qubits, dressed coherent state (dressed displaced vacuum), or
dressed squeezed vacuum. We have also shown that one can observe
quantum interference of states, transmitted in two-dimensional
lattices through various paths differing solely in the numbers of
intermediate sites of each path. This interference effect is
analogous to the signature-dependent effects in nuclear-structure
theories~\cite{Ring}.

\section*{ACKNOWLEDGEMENTS}

This work was supported in part by the NSA, LPS, ARO, NSF Grant
No. EIA-0130383, and JSPS-RFBR No. 06-02-91200. L.A.W. has been
supported by the Ikerbasque Foundation. A.M. acknowledges support
from the Polish LFPPI Network.


\end{document}